\documentclass[aps,10pt,superscriptaddress,twocolumn,showkeys,amsmath,amssymb]{revtex4-1}

\usepackage{graphicx}
\usepackage{bm}
\usepackage[breaklinks=true,colorlinks,citecolor=blue,linkcolor=blue,urlcolor=blue]{hyperref}
\usepackage{color}
\usepackage{amsmath}
\usepackage{datetime}
\usepackage{gensymb}
\usepackage[normalem]{ulem}

\begin{document}

\title{Programmable quantum Hall bisector: towards a novel \\ resistance standard for quantum metrology}
\author{Zahra Sadre Momtaz}
\affiliation{NEST, Istituto Nanoscienze--CNR and Scuola Normale Superiore, Piazza S. Silvestro 12,
I-56127 Pisa, Italy}
\author{Stefan Heun}
\affiliation{NEST, Istituto Nanoscienze--CNR and Scuola Normale Superiore, Piazza S. Silvestro 12,
I-56127 Pisa, Italy}
\author{Giorgio Biasiol}
\affiliation{IOM CNR, Laboratorio TASC, Area Science Park Basovizza, I-34149 Trieste, Italy}
\author{Stefano Roddaro }%
\email{stefano.roddaro@unipi.it}
\affiliation{NEST, Istituto Nanoscienze--CNR and Scuola Normale Superiore, Piazza S. Silvestro 12,
I-56127 Pisa, Italy}
\affiliation{Department of Physics ``E.Fermi'', Universit\`a di Pisa, Largo Pontecorvo 3, I-56127 Pisa, Italy}

\date{\today}

\begin{abstract}
We demonstrate a programmable quantum Hall circuit that implements a novel iterative voltage bisection scheme and allows obtaining any binary fraction $(k/2^n)$ of the fundamental resistance quantum $R_K/2=h/2e^2$. The circuit requires a number $n$ of bisection stages that only scales logarithmically with the precision of the fraction. The value of $k$ can be set to any integer between 1 and $2^n$ by proper gate configuration. The architecture exploits gate-controlled routing, mixing and equilibration of edge modes of robust quantum Hall states. The device does not contain {\em any} internal ohmic contact potentially leading to spurious voltage drops. Our scheme addresses key critical aspects of quantum Hall arrays of resistance standards, which are today widely studied and used to create custom calibration resistances. The approach is demonstrated in a proof-of-principle two-stage bisection circuit built on a high-mobility GaAs/AlGaAs heterostructure operating at a temperature of $260\,{\rm mK}$ and a magnetic field of $4.1\,{\rm T}$.
\end{abstract}

\maketitle

\section{\label{sec:intro} Introduction}

The integer quantum Hall (QH) effect~\cite{vonklitzing1980,sarma2008,chakraborty2013,cage1990} is one of the cornerstones of metrology and provides a precise experimental realization of the von Klitzing resistance quantum $R_{K} = h/e^2$~\cite{vonklitzing1985,vonklitzing1986,jeckelmann1997,jeckelmann2001,witt1998}, where $h$ is the Planck constant and $e$ the electron charge. Integer QH resistance is exactly quantized $R_H=R_K/i$, where $i$ is the integer number of filled Landau levels in the device, but calibration procedures would greatly benefit from the availability of generic quantum Hall resistance (QHR) standards, and decadic values (e.g. $1\,{\rm k\Omega}$, $10\,{\rm k\Omega}$ etc) in particular. This is typically achieved by using resistance ratio bridges, which are based -- in their most accurate and refined implementation -- on cryogenic current comparator bridges~\cite{witt1998,schopfer2007,poirier2009,williams2011,hernandez2014}. The technical complexity entailed by these methods has motivated the development of alternative strategies to produce new QHRs.

Advanced QHR can be obtained by creating devices that combine different quantized regions that are either: {\em (i)} {\em contiguous}, as in the case of a unique two-dimensional electron system (2DES) with a space-dependent filling factor; or {\em (ii)} {\em connected} by a set of low-resistance ohmic contacts and metallic connections. Devices combining multiple filling factors in the same electron system typically display complex non-local resistance effects~\cite{syphers1984,buttiker1988} and require the use of small filling factors since large mobility gaps are necessary to obtain a robust quantization~\cite{cage1990}. Because of these reasons, they have been subject to a limited investigation in the context of metrology. The second option has led to the creation of quantum Hall arrays of resistance standards (QHARSs), that integrate large networks of distinct identical Hall bars connected in series/parallel and can yield a rescaled resistance $(p/q)\times R_H$ where $p$ and $q$ are integer numbers~\cite{piquemal1999,poirier2002}. QHARSs provide a practical way to obtain a variety of QHRs, but they typically require a large number of elements and, despite the discovery of effective mitigation techniques~\cite{delahaye1993,delahaye1995,jeffery1995}, they critically depend on the quality of a large number of internal ohmic contacts traversed by a finite current flow. This obviously introduces internal stray voltage drops that hamper the precision of the device. In spite of the progress in the synthesis and modeling of innovative QHARSs~\cite{ortolano2011,ortolano2014,poirier2004, marzano2018} that maximize their simplicity and accuracy while minimizing the number of integrated elements, these limitations are still relevant.

\begin{figure*}[ht!]
\includegraphics[width=\textwidth]{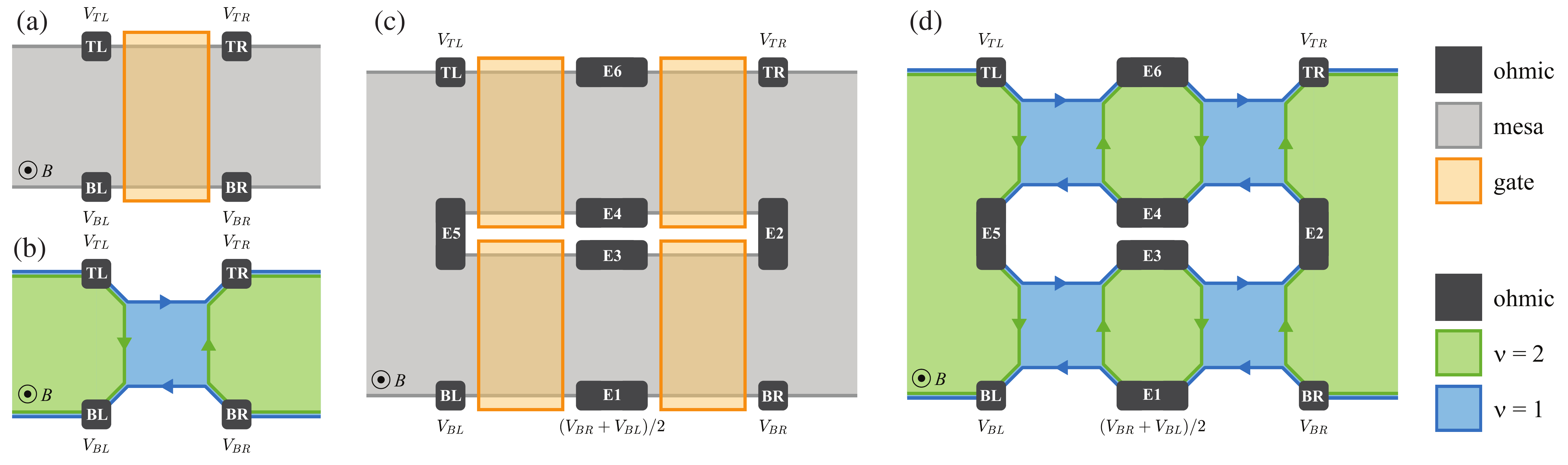}
\caption{\label{fig:fig1} (a,b) Edge mixing can be obtained by patterning a single barrier on an electron gas at filling factor $\nu=2$. The edge connectivity for the two Landau levels is visible in (b): due to edge equilibration, the top-right and bottom-left contacts both equilibrate at an average voltage $V_\mathrm{TR} = V_\mathrm{BL} = (V_\mathrm{TL} + V_\mathrm{BR}) / 2$. (c) The action of  a single mixer is fully equivalent to the double-parallel-double-serial (DPDS) mixer: this implies that any individual barrier can be replaced by a DPDS barrier without affecting the currents and potentials of the outer edge system. (d) The effect of equilibration in the DPDS mixer can be easily deduced from the edge connectivity scheme, as discussed in the main text.}
\end{figure*} 

The possibility of creating QHRs that include contiguous regions at different filling factors has been recently revived by the development of graphene-based resistance standards~\cite{giesbers2008,poirier2010,bounouh2003,tzalenchuk2010,janssen2013,novikov2016,park2020}. Indeed graphene -- beyond displaying sizably larger Landau gaps -- can combine $p$-type and $n$-type regions leading to non-standard edge configurations at their junction~\cite{matsuo2015,abanin2007,williams2007,woszczyna2011}. Recent studies have even highlighted how the local control of the filling factor by field-effect can be used to create {\em programmable} QHRs~\cite{hu2018}. Nevertheless, devices demonstrated so far still typically rely to some extent -- similarly to QHARS -- on internal resistive connections, and the number of elements they integrate does not scale favorably as a function of the precision of the required $(p/q)$ fraction. 

Here, we discuss and experimentally demonstrate a radically different QH architecture that allows obtaining an electrically-programmable QHRs that can produce {\em any} binary fraction $(k/2^n)$ of the resistance quantum $R_K/2$. Remarkably, the size of the circuit only grows linearly with the integer $n$, and thus logarithmically with the precision of the required fraction. Every integer $k$ value between $1$ and $2^n$ can be obtained with an easy-to-predict gate configuration and, regardless of the value of $n$, the circuit only requires using robust QH states at low filling factors. Finally, the device architecture does not contain any internal ohmic contact with a net current flow, and thus it is fundamentally less susceptible than conventional QHARSs to errors due to spurious contact resistances. We illustrate a proof-of-principle $n=2$ implementation based on a GaAs/AlGaAs heterostructure and report a relative accuracy of the output fractions of the order of $10^{-4}$, which is at the limit of the precision currently attainable with our measurement set-up.

The paper is organized as follows: in Section II we describe the key building block of the programmable QHRs, which consists in a voltage bisection scheme that is not affected by the non-local resistance effects that typically occur in QH circuits with a complex edge connectivity; in Section III we describe the structure of the full device; in Section IV we present the experimental results obtained on our proof-of-principle devices; and finally, in Section V, conclusions are drawn.

\section{Edge mixers}

The basic building block of the device is depicted in Figs.~\ref{fig:fig1}a,b: it consists of a field-effect transistor with a set of equilibration contacts and implements a QH edge-state mixer. The device is designed to operate at a filling factor $\nu=2$ in the ungated 2DES. The filling factor underneath the barrier region depends on the gate voltage: when the filling factor below the gate is set to $\nu=1$, the mixer will be said to be in an {\em ``active''} configuration; when the filling is $\nu=2$, the mixer will be said to be in a {\em ``neutral''} configuration. The full-depletion case is not relevant to the current paper and will not be taken into account. 

The operation of the mixer is straightforward. When the mixer is ``active'', only the outer edge channel is able to cross the barrier while the inner one is completely reflected, as sketched in Fig.~\ref{fig:fig1}b. Given the chirality in the figure (set by the sign of the magnetic field), the Landauer-B\"uttiker formalism~\cite{buttiker1988} implies that the two output voltages $V_\mathrm{TR}$ and $V_\mathrm{BL}$ equilibrate to the average value of the two input voltages $V_\mathrm{TL}$ and $V_\mathrm{BR}$, i.e.
\begin{equation}
V_\mathrm{BL} = V_\mathrm{TR} = (V_\mathrm{TL} + V_\mathrm{BR})/2.
\end{equation}
We note that the equilibration will naturally occur during the copropagation of edge modes~\cite{paradiso2011} and that it can be reasonably assumed to be extremely good for a macroscopic ohmic contact; this expectation likely deserves a proper experimental quantification, but required methods go beyond the level of accuracy reachable in the current paper. The crucial principle behind the voltage bisection scheme is the following: the single mixer in Fig.~\ref{fig:fig1}a is {\em electrically equivalent} to the combination of barriers and equilibration contacts visible in Fig.~\ref{fig:fig1}c,d which in the following will be named double-parallel-double-serial (DPDS) mixer. Here, ``{\em equivalent}'' means that any single barrier can be replaced with a DPDS barrier system without affecting in any way the voltages and the currents in the rest of the circuit. This is a basic consequence of the Landauer-B\"uttiker theory: as explained in detail in the following, once the input voltage $V_\mathrm{TL}$ and $V_\mathrm{BR}$ are set, the contacts $V_\mathrm{TR}$ and $V_\mathrm{BL}$ equilibrate at the same voltages as in the single barrier; this in turn ensures that all the input and output currents are exactly the same and thus that the two edge circuits are indistinguishable from the point of view of the rest of the QH circuit. The DPDS edge mixing scheme is fairly more complicated than a single mixer and indeed includes 6 new equilibration contacts E1-E6 (see Figs.~\ref{fig:fig1}c,d). Nevertheless, the behavior of the circuit can be deduced in a straightforward way from the following set of simple linear equations, one for each mixer:
\begin{equation}
   \left\{
	    \begin{array}{c}
         V_{\rm{E5}} = V_{\rm{E6}} = (V_\mathrm{TL}+V_{\rm{E4}})/2\\
         V_{\rm{E4}} = V_\mathrm{TR} = (V_{\rm{E6}}+V_{\rm{E2}})/2\\
         V_\mathrm{BL} = V_{\rm{E3}} = (V_{\rm{E5}}+V_{\rm{E1}})/2\\
         V_{\rm{E1}} = V_{\rm{E2}} = (V_{\rm{E3}}+V_\mathrm{BR})/2.
      \end{array}
	 \right.
\end{equation}

Simple substitutions lead again to equation (1), i.e. the circuit reproduces the behavior of the single mixer in Figs.~\ref{fig:fig1}a,b. Therefore, {\em any} single barrier can be replaced by a DPDS mixer without affecting the rest of the edge circuit. This is a crucial and non-obvious property of the circuital scheme we describe, since in most of QH circuits a local modification of the edge-channel connectivity differently leads to non-trivial non-local effects that can substantially change the global behavior of the device. It is useful to stress that the procedure can be iterated multiple times also on any of the newly introduced barriers in the DPDS mixer.

Furthermore, a new intermediate voltage $(V_\mathrm{BL}+V_\mathrm{BR})/2$ is available at the equilibration probe E1 located between BL and BR. As a consequence, any of the two gates at the bottom of Fig.~\ref{fig:fig1}c can be in turn replaced by a DPDS so to obtain a voltage which is intermediate between BL and E1, or between E1 and BL. This bisection scheme can be iteratively repeated to obtain any binary fractional value in the interval between the initial voltages $V_\mathrm{BL}$ and $V_\mathrm{BR}$. This procedure can be expected to lead to a rather complex multiply connected circuit architecture but, as discussed in the next paragraph, a simple reconfigurable version of this approach can be implemented.

\begin{figure}[ht!]
\includegraphics[width=0.85\columnwidth]{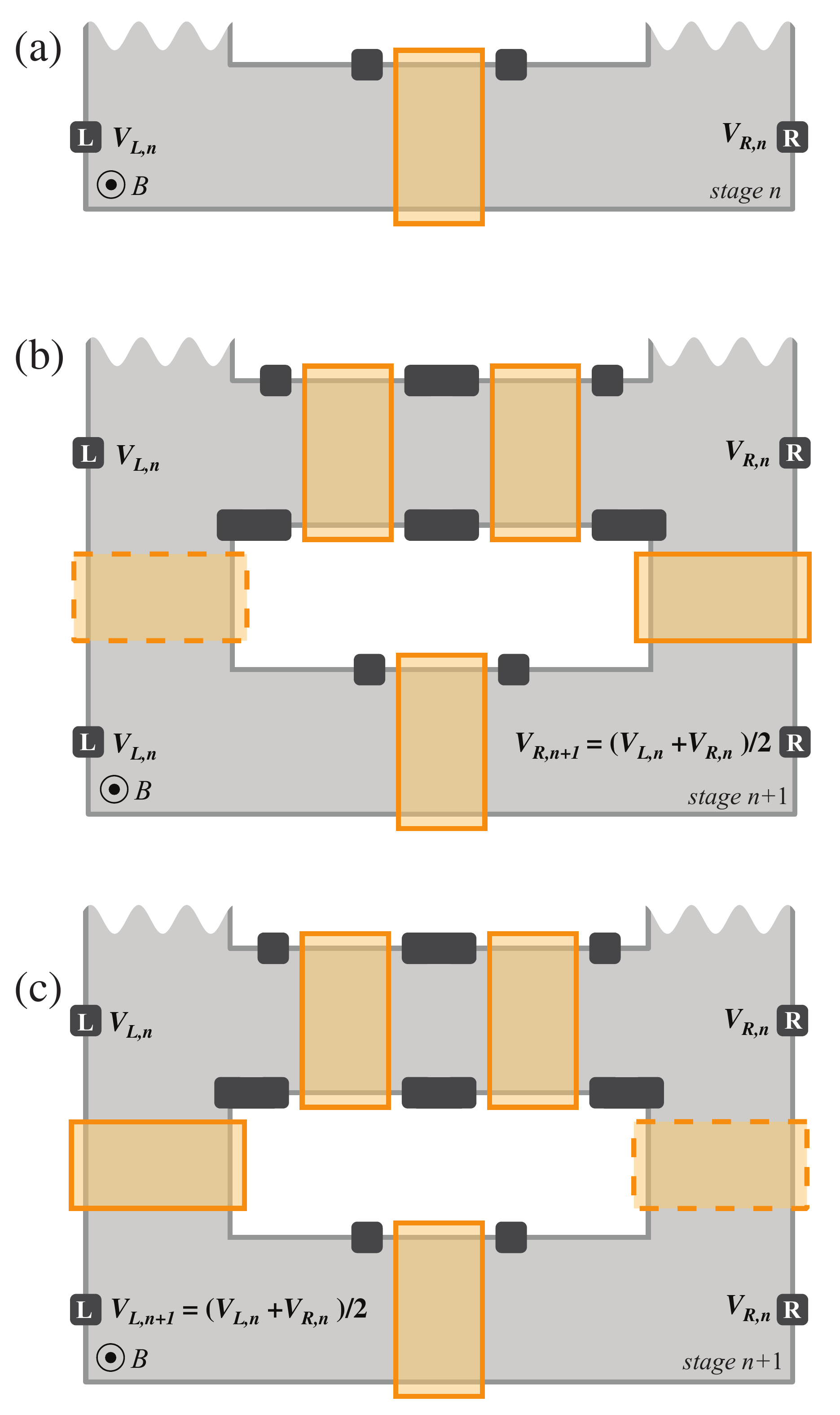}
\caption{\label{fig:fig2} (a) Single-barrier mixer with initial left and right voltages $V_\mathrm{L,n}$ and $V_\mathrm{R,n}$. (b,c) The barrier can be replaced by a DPDS mixer without affecting the rest of the circuit. The two complementary gate configurations (dashed gate is inactive) lead to reconfigurable output voltages $V_\mathrm{L,n+1}$ and $V_\mathrm{R,n+1}$. The circuit can be used to implement and iterative voltage bisection scheme, as discussed in the main text.}   
\end{figure} 

The individual bisection step described above can be obtained as a reconfigurable QH circuit allowing a more natural stacking of multiple bisection stages. The key idea is illustrated in Fig.~\ref{fig:fig2}: we report a simple mixer, whose behavior is identical to the topologically equivalent scheme in Fig.~\ref{fig:fig1}a; in the sketch, we highlight the BL and BR contacts and simply rename them as L and R. Since this circuit will be regarded as the $n^{th}$ bisection stage of a larger circuit, we name the voltages at the L and R contacts as $V_\mathrm{L,n}$ and $V_\mathrm{R,n}$, respectively. As in Fig.~\ref{fig:fig1}, the barrier can be replaced by a DPDS circuit but here we adopt a redundant gate scheme that allows a reconfiguration of the bisection stage. In Fig.~\ref{fig:fig2}b, the dashed gate on the left side is assumed to be in a ``neutral'' configuration, i.e. it has no effect on the edge system and can be neglected. All the other gates are assumed to be in an ``active'' configuration. In this case, which we will call configuration ``0'', we expect
\begin{align}
V_\mathrm{L,n+1} &= V_\mathrm{L,n} \nonumber\\
V_\mathrm{R,n+1} &= (V_\mathrm{L,n} + V_\mathrm{R,n})/2 = V_\mathrm{R,n} - \Delta V_\mathrm{n} / 2 \nonumber
\end{align}

\noindent where we define the $n^{th}$ voltage interval $\Delta V_\mathrm{n} = V_\mathrm{R,n}-V_\mathrm{L,n}$. In Fig.~\ref{fig:fig2}c, we depict the configuration ``1'' of the stage, where the gate on the right side is inactive, and the contacts equilibrate to 
\begin{align}
V_\mathrm{L,n+1} &= (V_\mathrm{L,n} + V_\mathrm{R,n})/2 = V_\mathrm{L,n} + \Delta V_\mathrm{n} / 2 \nonumber\\
V_\mathrm{R,n+1} &= V_\mathrm{R,n}. \nonumber
\end{align}

Note that in both configurations the new voltage interval is $\Delta V_\mathrm{n+1} = \Delta V_\mathrm{n} / 2 $. If we introduce the binary constant $c_\mathrm{n+1}$ indicating the configuration of the $(n+1)^{th}$ bisection stage $[c_\mathrm{n+1} = 0$ or $c_\mathrm{n+1} = 1]$, both results can be described by 
\begin{eqnarray}
   V_\mathrm{L,n+1} & = & V_\mathrm{L,n}+c_\mathrm{n+1}\cdot \Delta V_\mathrm{n+1} \\
   V_\mathrm{R,n+1} & = & V_\mathrm{L,n+1} + \Delta V_\mathrm{n+1}.
\end{eqnarray}

This method can be iteratively applied to the bottom barrier and allows obtaining any binary fractional value between two initial extremal voltages $V_\mathrm{L,0}$ and $V_\mathrm{R,0}$. In particular, after $n$ bisections we have
\begin{align}
V_\mathrm{L,n} &= V_\mathrm{L,0}+\sum_{i=1}^n c_i \cdot \frac{\Delta V_0}{2^i} \\
&= V_\mathrm{L,0}+\sum_{i=1}^n c_i2^{n-i}\cdot\frac{\Delta V_0}{2^n} \\
&= V_\mathrm{L,0}+(k-1)\cdot\frac{\Delta V_0}{2^n}.
\end{align}

\noindent where we introduced an integer $k$ that runs, depending on the binary values $c_i$, from $1$ to $2^n$. Finally we can also write $V_\mathrm{R,n}=V_\mathrm{L,0}+(k/2^n)\cdot\Delta V_0$ and notice that the configuration of the bisection stages is directly connected to the binary representation of $k-1$. Importantly, the number $n$ of bisection stages grows only logarithmically with the magnitude of the denominator $2^n$. 

To conclude the illustration of the device working principle, it is important to note that the bisection scheme described here does not introduce any spurious potential drop due to internal connections, such as typically occurs in the case of conventional QHARS and similar QH schemes. As such, the described approach implements a pure four-wire scheme that is not fundamentally limited by the contact quality. This bodes well for potential applications in the context of metrology.

\section{\label{sec:architecture} Circuit architecture}

\begin{figure}[ht!]
\includegraphics[width=0.95\columnwidth]{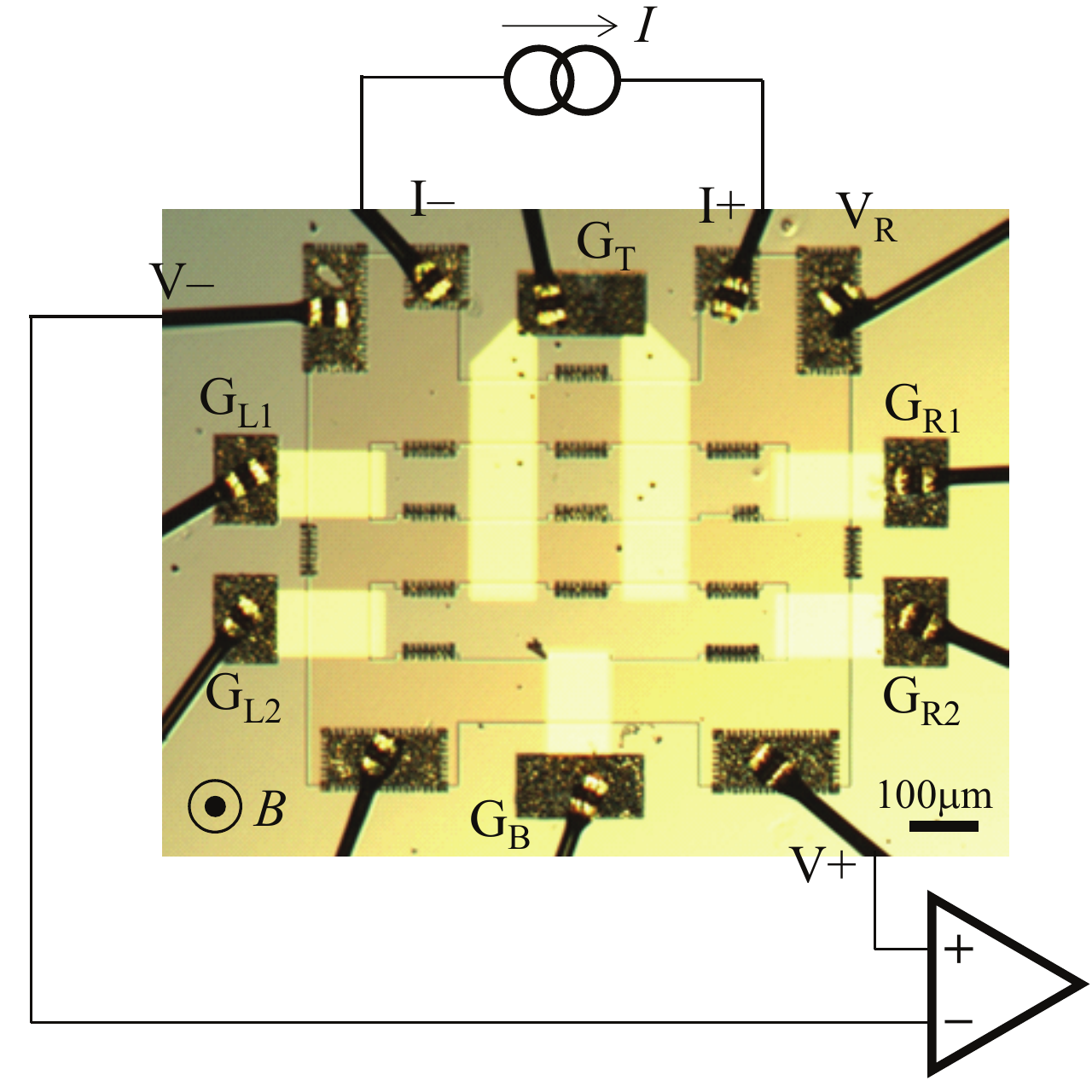}
\caption{\label{fig:fig3} Optical micrograph of one of the studied proof-of-principle circuits. The device implements a two-stage bisector that can produce the following quantized resistances: $h/8e^2$, $h/4e^2$, $3h/8e^2$ and $h/2e^2$. The value can be tuned by choosing a suitable configuration of the lateral control gates of the bisection stages.}
\end{figure}

The staged bisector described in the previous section was implemented using a high mobility  GaAs/AlGaAs single-heterojucton 2DES located at $100\,{\rm nm}$ below the surface. The architecture of the final device is illustrated in the optical picture of Fig.~\ref{fig:fig3} and implements a two-stage demonstrator of the iterative bisection scheme of Fig.~\ref{fig:fig2}; a further sketch of the device structure is reported in the Supplementary Information. The ohmic contact pads are fabricated by UV-lithography, followed by thermal evaporation of a Ni/GeAu/Ni/Au ($100/10/200/10\,{\rm nm}$) multilayer and standard rapid thermal annealing. The mesa was defined by H$_3$PO$_4$ wet etching in a second UV-lithographic step. Finally, the gate electrode layer was obtained in a third UV-lithography step followed by thermal evaporation of Ti/Au ($10/50\,{\rm nm}$).

The device is designed to operate at $\nu=2$ in the gate-free regions of the heterostructure. In addition, it contains a set of central gates that need to be configured to the ``active'' state, i.e. at $\nu=1$. The gates are controlled by the bottom pad $\rm{G_B}$ and by a common top pad $\rm{G_T}$. Two sets of side electrodes ${\rm G_{Ln}}$ and ${\rm G_{Rn}}$ control the status of the two bisection stages. All inner equilibration contacts are fabricated so as to expose a large interface to the 2DES and ensure full edge equilibration. All barriers and channels have a large size of about $100\,{\rm \mu m}$ to minimize spurious backscattering due to disorder. With respect to the design in Fig.~\ref{fig:fig2}, the top part of the device implements four additional ohmic contacts to drive the circuit with a given current $I$. Differential voltage measurements are using contacts $V_\mathrm{L,0}$ and $V_\mathrm{R,n=2}$. At $\nu=2$, for a given drive current $I$, the Landauer-B\"uttiker formalism implies $V_\mathrm{R,0} = hI/2e^2$ and thus $\Delta V_0 = hI/2e^2$. This leads to 
\begin{align}
\frac{V_{\rm R2}-V_{\rm L0}}{I} = \left(\frac{k}{2^n}\right)\cdot\frac{h}{2e^2}
\end{align}
and by properly biasing the control gates we thus expect to obtain the fractional resistance quanta $R=h/8e^2$ (bisection configuration ``00''), $h/4e^2$ (``01''), $3h/8e^2$ (``10'') and $h/2e^2$ (``11''). 

\begin{figure}[ht!]
\includegraphics[width=0.98\columnwidth]{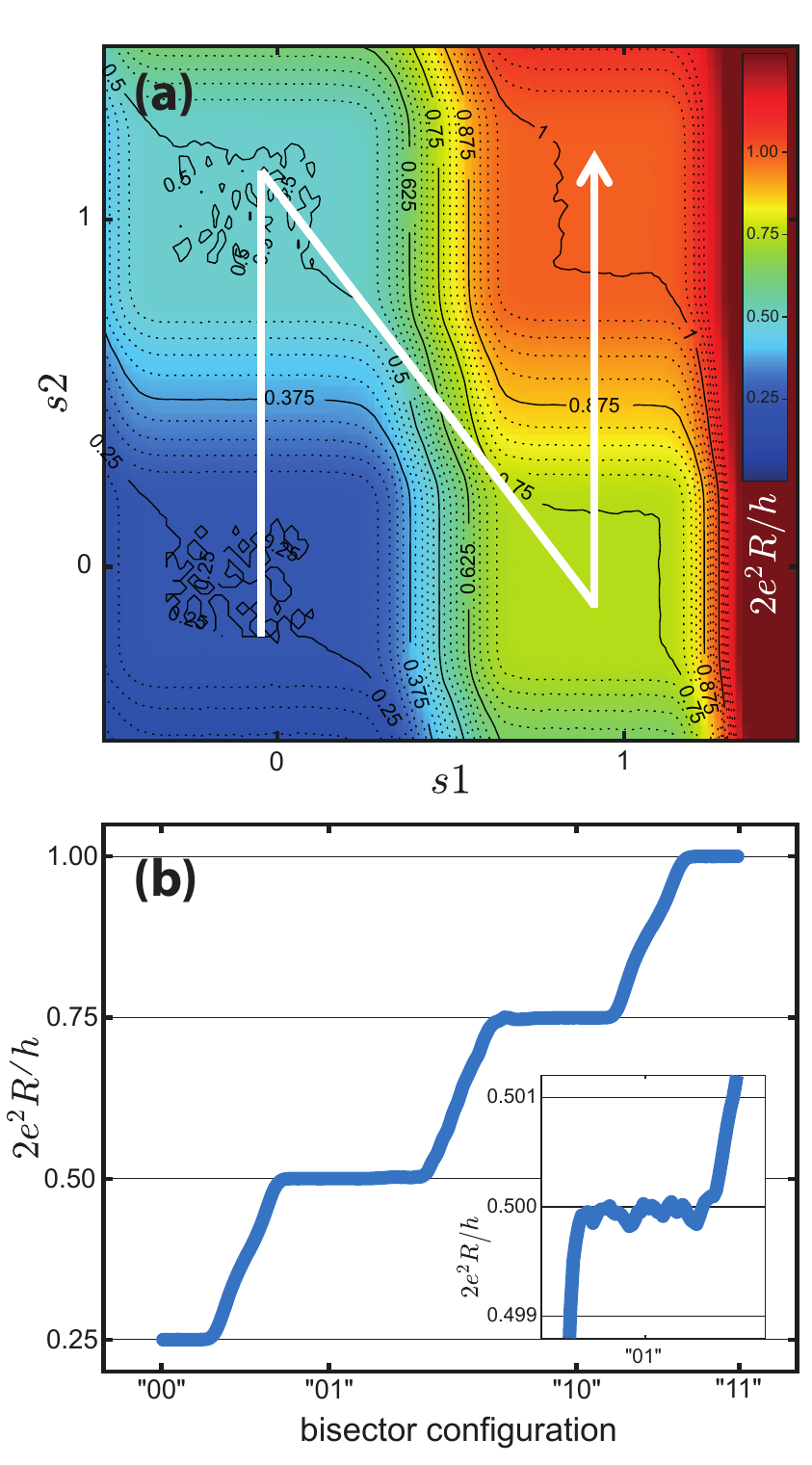}
\caption{\label{fig:fig4} (a) Experimental demonstration of a two-stage bisection circuit implementing four different quantized resistances depending on the two control parameters $s_1$ and $s_2$. Four plateus are visible at the fractions $2e^2R/h=1/4$, $1/2$, $3/4$ and $1$. (b) Cross-section along the trajectory indicated by the white line in (a). The precision of the quantization can be assessed from the inset showing the $0.5$ plateau, and from the statistical analysis reported in Fig.~\ref{fig:fig5}. }
\end{figure}

\begin{figure*}[ht!]
\begin{center}
\includegraphics[width=0.99\textwidth]{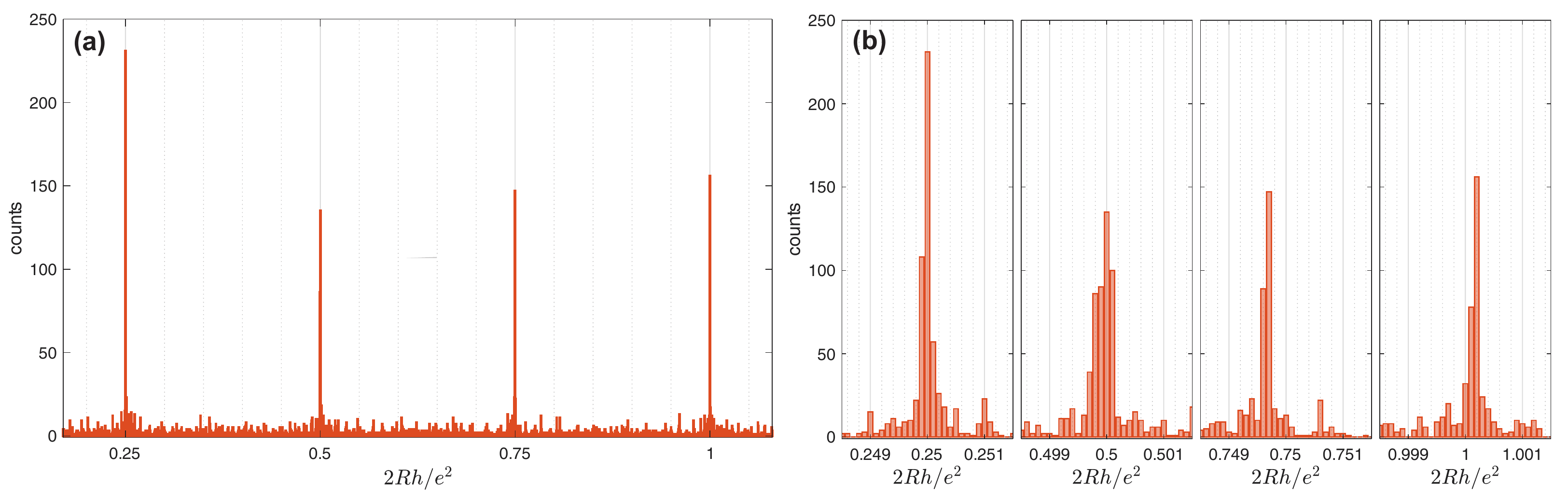}
\caption{\label{fig:fig5} (a) Histogram of the whole dataset reported in Fig.~\ref{fig:fig4}. No clipping was performed to select the quantization plateaus. A sharp concentration of values is observed in correspondence to the fractions $1/4$, $1/2$, $3/4$ and $1$. (b) The precision of the quantization can be better assessed in the zoomed plots. The quantized plateaus are consistent within an error of the order of $10^{-4}$, which is at the limit of the current measurement setup.}
\end{center}
\end{figure*}

Magneto-transport experiments were performed using a phase-locked technique at $17\,{\rm Hz}$ and at a temperature of $T = 260\,{\rm mK}$. The ideal current source indicated in Fig.~3 was obtained using the lock-in AC source and a $10\,{\rm M\Omega}$ bias resistor. Since metallic gates naturally induce some depletion and backscattering even when grounded, it is important to calibrate the device so as to find the finite (positive) bias voltage that neutralize the action of the gate electrodes. As discussed in the Supplementary Information, this can be experimentally achieved based on the value of the ``pseudo-longitudinal'' resistance measurement using the voltage probe pair $V_\mathrm{R,0}$ and  $V_\mathrm{R,2}$. A gate bias of at least $\approx +0.1\,{\rm V}$ was found to be necessary to obtain dissipationless transport in the QH regime. The basic transport parameters of the electron system were directly measured on the final devices with all gates biased at their neutral configuration. Based on QH evidences and on a numerical simulation of the actual device geometry (see Supplementary Information), we estimate a carrier density $n = 1.9\times 10^{11}\,{\rm cm^{-2}}$ and electron mobility $\mu\approx0.5\times10^6\,{\rm cm^2/Vs}$. A robust $\nu=2$ QH state is observed in the range $B=3.8-4.4\,{\rm T}$. Measurements reported in the next section were performed at $B=4.1\,{\rm T}$. 

\section{\label{sec:operation} Device operation}

The operation of the bisection circuit is demonstrated in Figs.~\ref{fig:fig4} and \ref{fig:fig5}. The first step consists in a precise calibration of all mixers, i.e. the gate voltages corresponding to the ``neutral'' versus ``active'' state have to be determined. The procedure is described in the Supplementary Information and involves the measurement of the voltage drops along the edge system versus gate voltage: clear plateaus can be observed as a function of the gate voltage and allow a clear identification of the bias necessary to obtain $\nu=2$ or $\nu=1$ in the gated region. The result of the calibration is reported in the Tab.~1.

\begin{table}
\begin{center}
\begin{tabular}{c|c|c|c}
Gate & stage & ``neutral'' state & ``active'' state \\
\hline
$\rm{G_{R1}}$ & 1 & $+0.120\,{\rm V}$ & $-0.025\,{\rm V}$\\
$\rm{G_{L1}}$ & 1 & $+0.110\,{\rm V}$ & $-0.065\,{\rm V}$\\
\hline
$\rm{G_{L2}}$ & 2 & $+0.100\,{\rm V}$ & $-0.050\,{\rm V}$\\
$\rm{G_{R2}}$ & 2 & $+0.200\,{\rm V}$ & $+0.060\,{\rm V}$\\
\end{tabular}
\caption{Calibration bias values for the mixers.}
\end{center}
\end{table}

Each configuration is typically found to be stable over a gate voltage interval of few tens of milliVolts. The device was operated by varying the gate voltages in a linear combination between the ``neutral'' and ``active'' states, using two continuous sweep parameters $s_1$ and $s_2$ for the two bisection stages, so that the ``active'' and ``neutral'' configurations for the bisection stage $i$ are obtained for $s_i=1$ and $s_i=0$, respectively. In the reported datasets we used
\begin{eqnarray}
   V_\mathrm{G_{R1}} & = & +0.120-0.145s_1 \\
   V_\mathrm{G_{L1}} & = & -0.065+0.175s_1 \\
   V_\mathrm{G_{R2}} & = & +0.200-0.140s_2 \\
   V_\mathrm{G_{L2}} & = & -0.050+0.150s_2
\end{eqnarray}

The resulting four-wire resistance $R$ as a function of $s_1$ and $s_2$ in the $[-0.5,+1.5]$ interval is reported in Fig.~\ref{fig:fig4} and contains clear plateaus at the predicted fractional resistance quanta, as visible in the contour plot in Fig.~\ref{fig:fig4}a. In the dataset, we report the measured resistance values obtained by using the $h/8e^2$ plateaus as a reference for the lock-in calibration. Using this procedure, the {\em relative} quantization values are found to be consistent within a precision on the order of few parts per $10^4$, thus demonstrating the working principle illustrated in the previous sections. Further details concerning this operational choice is reported in the Supplementary Information. 

The values of the different resistance plateaus are highlighted in the cross-sectional view in Fig.~\ref{fig:fig4}b, obtained by slicing the full dataset along the white line visible in Fig.~\ref{fig:fig4}a. In order to quantify the quality of the quantized plateaus, we created an histogram of the full set of measured $R$ values (see Fig.~\ref{fig:fig5}, no selection was done on the range of control parameters $s1$ and $s2$): very sharp peaks (full width half maximum $\approx 2\times 10^{-4}$) are observed in correspondence to the predicted fractions $1/4$, $1/2$, $3/4$ and $1$. The relative precision of the quantized values is very good and plateau ratios are exact within a precision well below $10^{-3}$. Further discussion on the attained precision is reported in the Supplementary Information.

For the sake of the possible application of the circuit as a resistance standard, it is worth to highlight few additional facts. As already mentioned, the circuit does not contain any internal contacts, thus it implements a true four-wire measurement scheme that is not affected by the quality of any of the contacts. We also highlight that the input impedance of the circuit does {\em not} change with the bisector configuration, owing to the equivalence discussed in the previous sections. Differently, the output resistance of the circuit is not negligible and any measurement performed with a finite input resistance set-up will affect the resistance value in a non-trivial way, which depends on the specific bisector configuration. In order to increase the precision it is thus important to introduce an high-impedance voltage preamplifier. Finally, we highlight that possible unideal effects might originate from the capacitive and DC coupling of the 2DES to the various gates included in the device architecture. The investigation of these effects is beyond the scope of the current paper.

To conclude, we would also like to highlight that of course many addition configurations could be obtained, beyond the ones visible in Fig.~\ref{fig:fig4}, by setting the bisection stages to configurations different from the states ``0'' and ``1''. For instance, both barriers can be set to be ``active'' or both ``neutral'', and in principle for an $n$ stage circuit there are $4^n$ different gate configurations, i.e. many more than those explored here. The resulting quantized resistance is less trivial to derive, and no unique formula was obtained so far. In principle these configurations could lead to further useful QHR values. However it is important to note that such additional configurations will display a different input impedance and possibly be associated with further systematic errors.

\section{Conclusion}
  
We discussed and demonstrated a novel edge mixing and equilibration architecture implementing a programmable QHRs. The circuit yields a generic binary fraction of the fundamental quantum $(R_k/2)$. The circuit represent a fundamental improvement of QHARSs: internal ohmic contacts with a net current flow are removed; the complexity of the circuit network only scales logarithmically as a function of the precision of the required conversion fraction; the circuit is reconfigurable using an easy-to-predict gating configuration. Additional studies will be required to test the viability of the approach at the levels of precision required by metrological applications, including in particular a quantification of the precision of edge equilibration at a macroscopic ($\approx 100\,{\rm \mu m}$) voltage probe and the impact of output impedance in association with finite impedance voltage measurement systems. No fundamental limitation to the method has been identified at the present time of the study.

\bibliography{QHrefs}

\newpage
\begin{center}
{\Large\bf Supporting Information:}
\end{center}

\setcounter{section}{0}
\section{Device structure and stage scaling}

\begin{figure}[ht!]
\includegraphics[width=0.7\columnwidth]{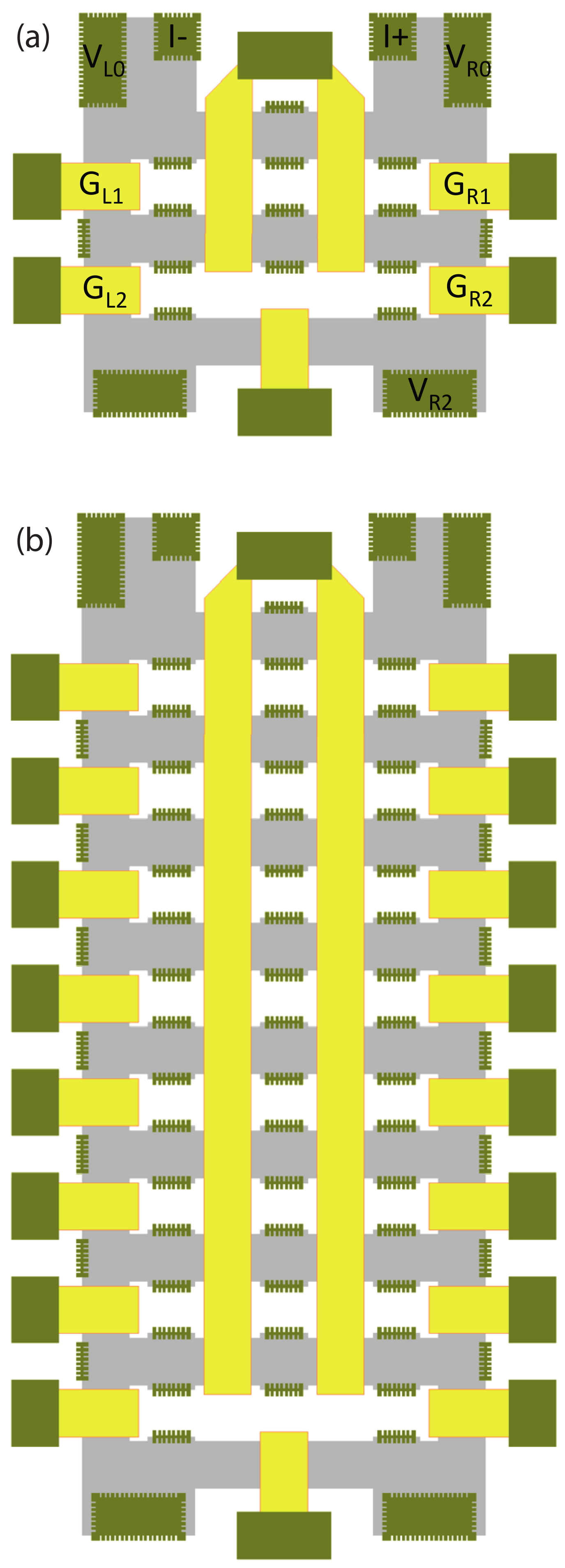}
\caption{\label{fig:fig-s1} (a) Sketch of the two-stage bisection circuit demonstrated in the main text. (b) Sketch of an eight-stage bisection circuit.}
\end{figure}

The procedure described in the main text can be iterated to obtain a QH circuit with a large number of stages, even if only up to two bisection stages -- sketch reported in Fig.~\ref{fig:fig-s1}a -- have been physically demonstrated so far. For the sake of clarity, in Fig.~\ref{fig:fig-s1}b we report the full schematics of an eight-stage bisection circuit, which can reproduce any resistance between $0$ and $h/2e^2\approx 12.9\,{\rm k\Omega}$ with a relative resolution $1:2^{8}=1:256$, corresponding to steps of resistance $R_K/512\approx50\,{\rm \Omega}$. Smaller steps can be obtained by adding further bisection stages, using a circuit whose complexity scales logarithmically with the target resolution.

\begin{figure*}[ht!]
\includegraphics[width=0.95\textwidth]{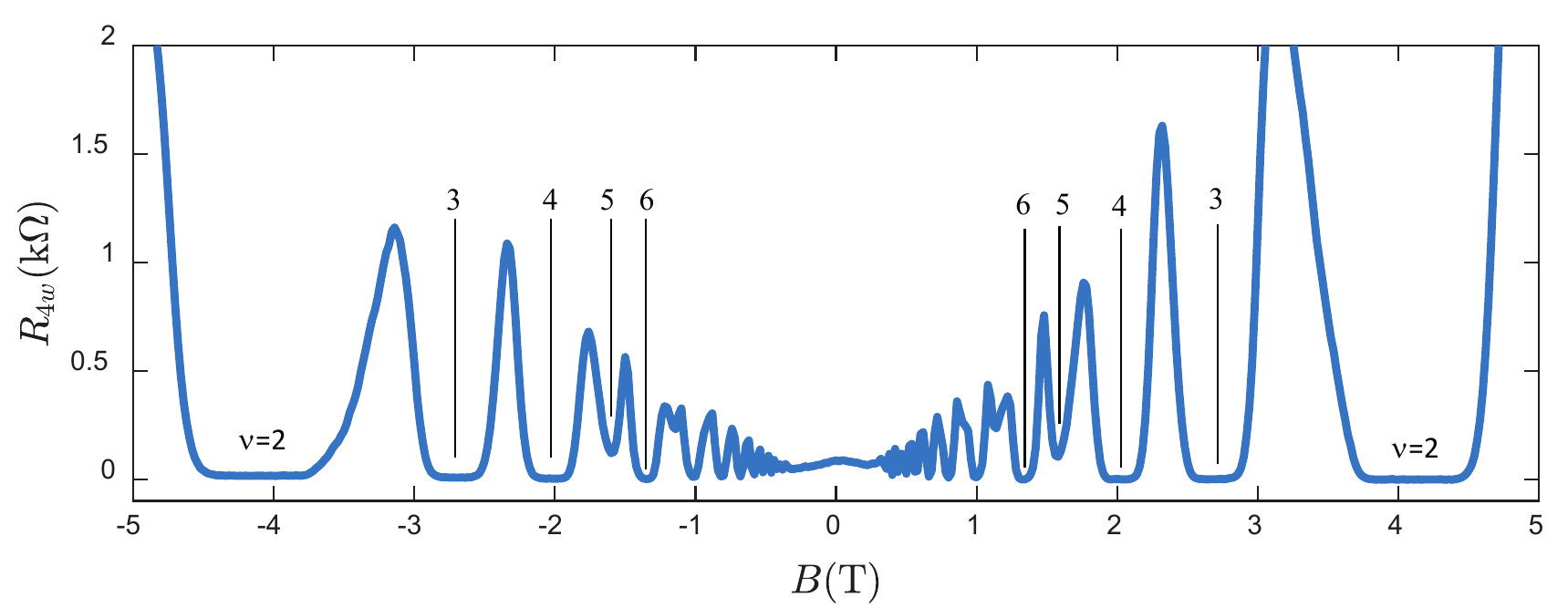}
\caption{\label{fig:fig-s2} Schubnikov-de Haas oscillations for the bisector circuit in a ``pseudo-longitudinal'' resistance configuration (see Fig.~\ref{fig:fig-s3}) where a suitable positive bias was applied to the gates to compensate the intrinsic depletion they induce in the electron system. Dissipationless transport is achieved for all the main QH states. A zero-field resistance of about $90\,{\rm\Omega}$ is obtained and used to estimate the average mobility of the 2D system.}
\end{figure*}

\section{Measurement set-up}

Measurements were performed using a conventional phase-locked technique: current biasing was achieved using a resistive load of $10\,{\rm M\Omega}$ and a nominal voltage excitation of $100\,{\rm mV}$, leading to a nominal current bias of $10\,{\rm nA}$. SR830 lock-in amplifiers by Stanford Research were used to measure currents and differential voltage drops across different pairs of contacts. Measurements have been carried out in a helium cryostat operating at $260\,{\rm mK}$. 

As reported in the main text, datasets in Fig.~4 and 5 were finely calibrated using the $h/8e^2$ plateau (corresponding to the $1/4$ fractional configuration of the bisection circuit) as a reference. Absolute resistance values can be calculated using the measured current as well, but simultaneous current and voltage measurements could not be performed: given the non-negligible modification of the circuit due to the finite input resistance of the lock-in preamplifiers and given the intrinsic calibration limits of the instrument, at present a relative comparison of the plateaus resulted to be the best procedure to quantify the precision of our bisection circuit. In this respect, it is important to consider the effect of the input and output resistances ($R_i$ and $R_o$) of the QH circuit. We note in particular that $R_i$, by design, is expected to be independent from the configuration of the bisection circuit and to correspond to a universal value $R_i=h/e^2$. This is a trivial consequence of the equivalence principle illustrated in the main text: the full bisection circuit -- from the point of view of the $I_+$ and $I_-$ current leads -- is always equivalent to a simple barrier at $\nu=1$.  The experimental behavior of the current in our non-ideal biasing scheme confirmed such expectation. Differently, the value of $R_o$ is not universal, even if we cannot provide at present a general expression for $R_o$ as a function of the circuit configuration. The output resistance was thus numerically calculated by solving the node equations for selected circuits with a few bisection stages (in particular, two stages) and $R_o\le 2h/e^2$ was always obtained, with minor variations $\Delta R_o$ between the different configurations of the bisection circuit. In particular, for the specific two-stage circuit we used to demonstrate our concept circuit, we calculated 
\begin{align}
R_{o,00}&=(27/32)R_K\approx21.78\,{\rm k\Omega},\\
R_{o,01}&=(28/32)R_K\approx22.59\,{\rm k\Omega},\\
R_{o,10}&=(35/32)R_K\approx28.23\,{\rm k\Omega},\\
R_{o,11}&=(32/32)R_K\approx25.81\,{\rm k\Omega},
\end{align}
with a maximum variation $\Delta R_o$ of $k\Omega$ between the different configurations. Since our measurements were performed with a standard lock-in with a nominal input resistance $R_{\rm LI}=10\,{\rm M\Omega}$, variations in the quantized values on the order of $\Delta R_o/R_{\rm LI}$, falling in the few $10^{-4}$ range, are to be expected. In this respect, we note that the minor deviations reported in Figs.~4 and 5 are partially consistent with the expected values of $R_o$ in the different configurations, in particular: (i) given $R_{o,00}$ is very similar to $R_{o,01}$ ($\Delta R_o<1\,{\rm k\Omega}$), the precise relative quantization value of ``01'' is non surprising; (ii) the lower relative quantization value measured for ``10'' is consistent with the larger $R_{o,10}$, leading to a larger voltage partitioning effect at the input of the lock-in preamplifier; (iii) based on output resistances above, a larger error should instead be expected for the relative quantization value of configuration ``11''. Besides these specific consistency checks, we note however that issues with the intrinsic linearity of the lock-in or further effects that have not been identified yet could not be ruled out at the moment. In conclusion, while the perspectives of the circuit for metrology applications are surely promising, their exact assessment goes beyond the scope of the current proof-of-principle paper and we have not attempted to further investigate the ultimate precision of our bisection circuit.

\section{Estimate of the electron mobility}

\begin{figure}[ht!]
\includegraphics[width=0.85\columnwidth]{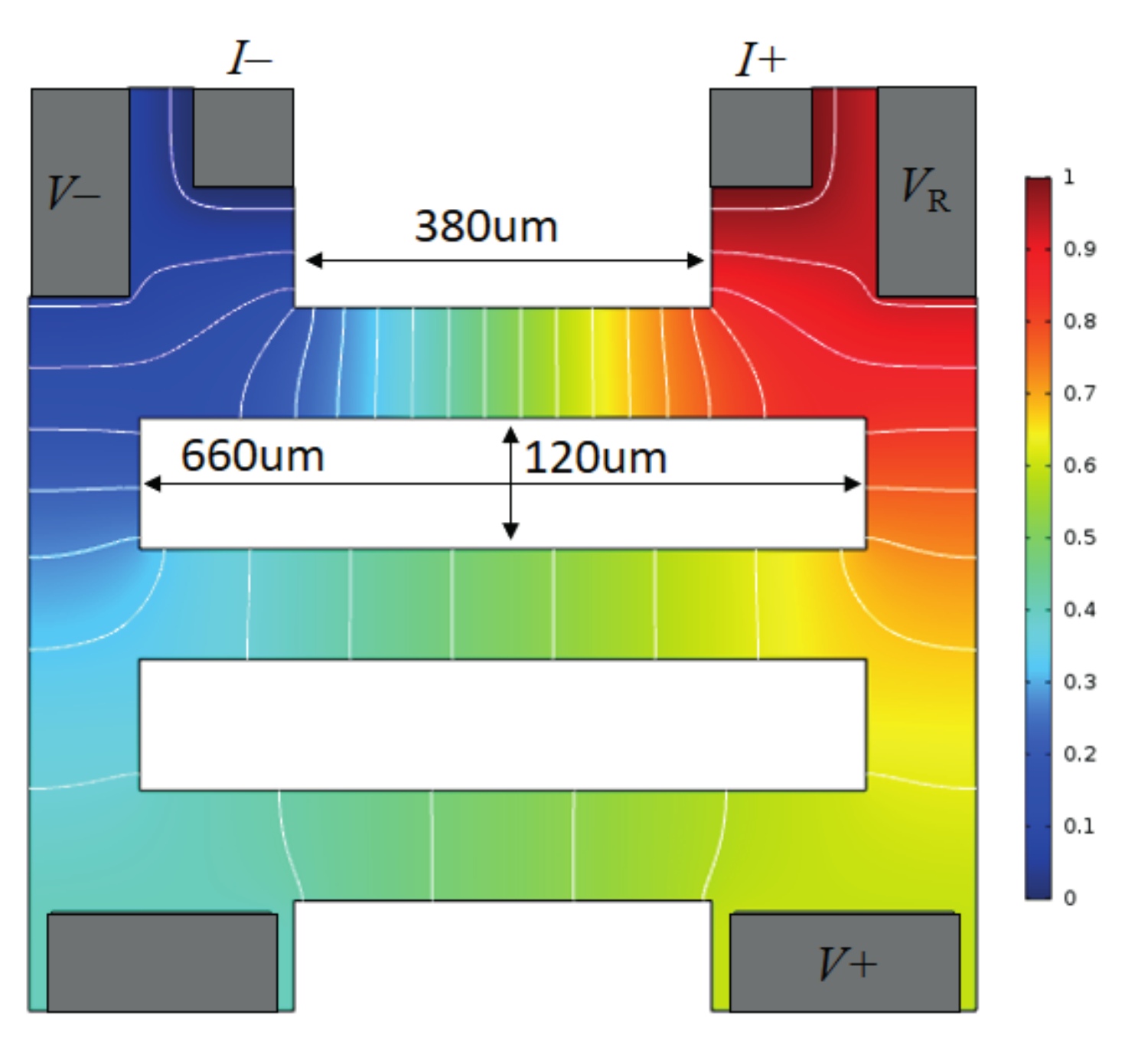}
\caption{\label{fig:fig-s3} The current flow in the device was simulated using a FEM model assuming a perfect neutralization of the gated region and neglecting the inner equilibration contacts. Based on magnetotrasnport measurements in the ``pseudo-longitudinal'' 4-wire configuration, we estimate an average electron mobility $\mu\approx 0.5\times10^6\,{\rm cm^2/Vs}$. In the simulation current is injected using the $I_\pm$ contacts while the four-wire resistance is calculated between $V_{\rm R}$ and $V_+$ (the so-called ``pseudo-longitudinal'' configuration cited in the main text).}
\end{figure}

The electron mobility in the complete device was estimated using a finite element calculation using MATLAB, starting from the known mesa geometry and based on magneto-transport data obtained while keeping all the gates at a positive gate voltage compensating their intrinsic depletion effect on the electron system (see Fig.~\ref{fig:fig-s2}). In the simulation reported in Fig.~\ref{fig:fig-s3}, the geometry depicted corresponds to the one actually implemented in the experiment. Gray rectangles represent the ohmic contacts and are set as fixed ($I_\pm$ contacts) or floating (all other) potential boundary conditions; on the rest of the boundary the ${\bf J} \cdot {\hat{\bf n}} = 0 $ condition is imposed. A voltage $V$ is applied to the contact $I_+$ while $I_-$ is kept grounded: the resulting current density is calculated to be equal to $I=0.23978\cdot\sigma V$, where $\sigma$ is the 2DES conductivity. The voltage drop between contacts $V_{\rm R}$ and $V_+$ is found to be equal to $0.338\cdot V$, leading to a four-wire resistance $R\approx1.42/\sigma$. The actual measured value $R\approx 90\,{\rm \Omega}$ (see zero field intercept in Fig.~\ref{fig:fig-s2}) leads to an estimated $\sigma = 0.0157\,{\rm S/\square}$. Using $n=1.9\times10^{11}\,{\rm cm^{-2}}$, this corresponds to $\mu\approx 0.5\times 10^{6}\,{\rm cm^2/Vs}$. The estimated average mobility is lower by a factor $2-3$ with respect to the nominal value obtained from the pristine wafer, which is not particularly surprising given all the processing steps and, in particular, the number of gated regions present in the final device.

\section{Gate calibration procedure}

The calibration procedure for the various gates consisted of a sequence of steps. First of all, the metallic gate electrodes are found to partially deplete the underlying electron system. This can be easily seen by looking at magneto-transport data using the so-called ``pseudo-longitudinal'' configuration (voltage drop between $V_\mathrm{R0}$ and $V_\mathrm{R2})$. When gates are grounded and the electron system is in the QH regime, one would ideally expect $V_\mathrm{R0}=V_\mathrm{R2}$. This is not what is observed in practice and a finite backscattering is observed at every value of the magnetic field $B$. True dissipationless transport at integer $\nu$ values (Fig.~S2) can only be obtained if a {\em positive} bias is applied to the gates so that a sufficiently uniform carrier density is re-established throughout the device. In particular, a voltage of about $+0.2\,{\rm V}$ was found to be necessary to reach such a condition for all the gates. The observation of clean Shubnikov-de Haas oscillation also allowed to precisely determine the best working condition to achieve $\nu=2$ in the ungated regions of the device. In the specific case reported in the paper, measurements were performed at a magnetic field of $4.1\,{\rm T}$.

In a second calibration step, central gates were biased to their correct operation point. The required bias was again determined by looking at the voltage drop between $V_\mathrm{R0}$ and $V_\mathrm{R2}$, which achieves a finite quantized value when the central gates are set to $\nu=1$. In the current layout, four out of five central gates were connected to a single bonding pad, so all central gates had to be biased to the same voltage. Given the relatively small size of the device, resistance data indicate that $\nu=1$ could be achieved simultaneously under all central barriers. We note that larger devices might require addressing individual barriers, if the homogeneity of the carrier density and/or of the processing is not ideal. 

\begin{figure}[ht!]
\includegraphics[width=0.92\columnwidth]{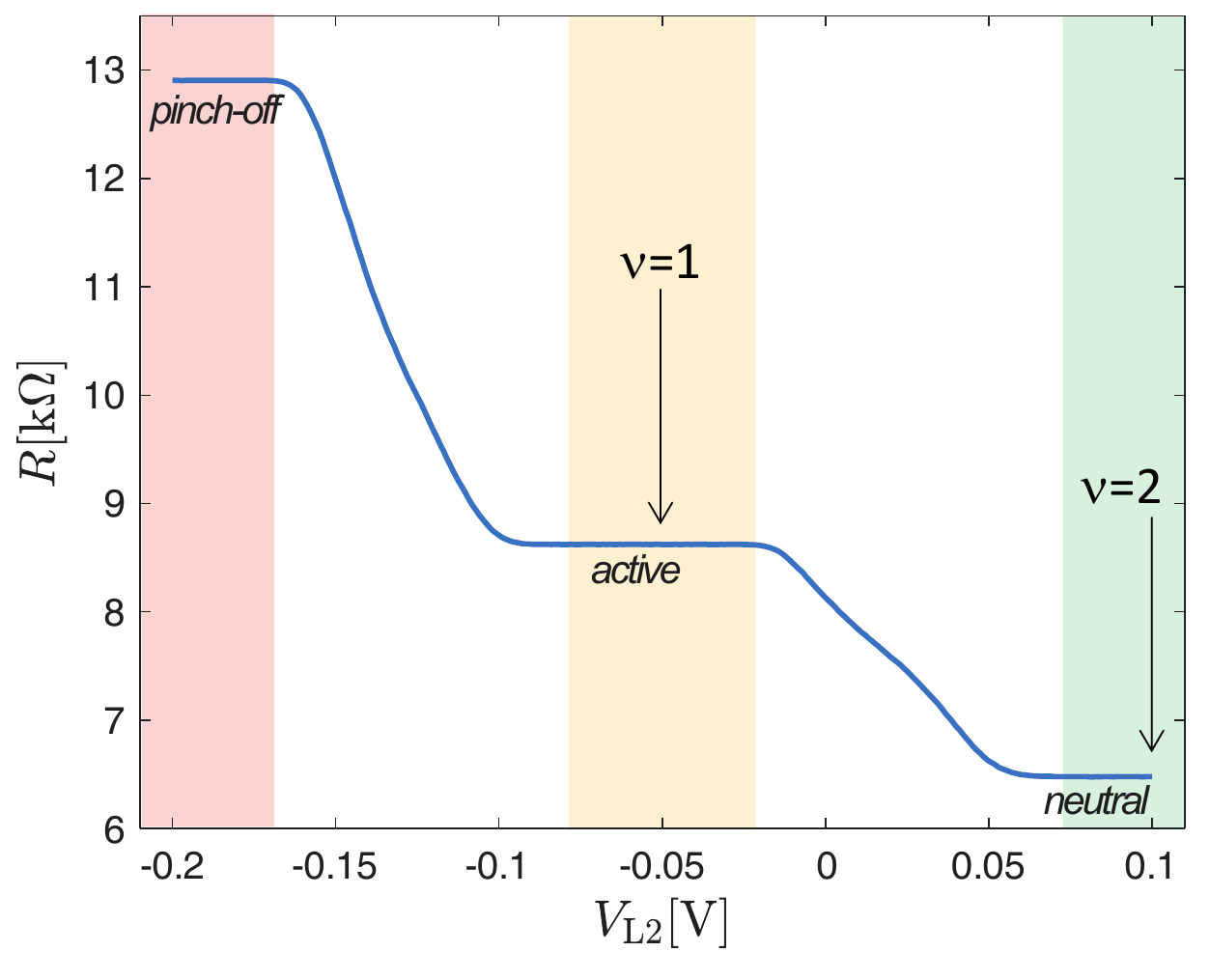}
\caption{\label{fig:fig-s4}Calibration of mixer L2. All the other lateral gates are kept at neutralizing voltage. The sweep of $V_{\rm L2}$ allows to identify a safe range for setting $\nu=2$ and $\nu=1$ under the field effect barrier.}
\end{figure}

In a third calibration step, once central gates are correctly configured, each of the lateral gates is sequentially calibrated so to identify the correct bias ranges to obtain $\nu=2$, $\nu=1$ and the pinch-off $\nu=0$ under the field-effect barrier. Upon biasing the gate, four wire resistances in the device display a stepwise increase with clear plateaus that allow a precise calibration. The results for gate $\mathrm{G_{L2}}$ are shown in Fig.~\ref{fig:fig-s4}. In the figure, we report the four wire resistance obtained by biasing the device through $I_\pm$ and recording the voltage drop on the pair $V_\mathrm{L0}$ and $V_\mathrm{R2}$. Depending on the specific barrier, different voltage probes can have a different sensitivity on the gate configuration and can be selected accordingly. Based on the reported data, we identify the bias for the ``active'' configuration as $V_{\rm L2}=-0.05\,{\rm V}$, and $V_{\rm L2}=+0.10\,{\rm V}$ has to be used to put the mixer in the ``neutral'' condition. A similar procedure was performed for $\mathrm{G_{L1}}$, $\mathrm{G_{R1}}$ and $\mathrm{G_{R2}}$.

\end{document}